\documentclass[draft]{agujournal2019}
\usepackage{url} %this package should fix any errors with URLs in refs.
\usepackage{lineno}
\usepackage{soul}
\usepackage{amsmath}
\usepackage{array}
\usepackage{appendix}
\usepackage{float}
\newcommand{\citet}[1]{\citeauthor{#1} \shortcite{#1}}

\drafttrue

\journalname{}
\makeatletter
\renewcommand{\@oddhead}{}
\let\@evenhead\@oddhead
\makeatother

\begin{document}

%%%%%%%%%%%%%%%%%%%%%%%%%%%%%%%%%%%%%%%%%%%%%%%
%  TITLE
%
% (A title should be specific, informative, and brief. Use
% abbreviations only if they are defined in the abstract. Titles that
% start with general keywords then specific terms are optimized in
% searches)
%
%%%%%%%%%%%%%%%%%%%%%%%%%%%%%%%%%%%%%%%%%%%%%%%

\title{How Do AI Climate Models Respond to Warming Across Climate Zones?}

%%%%%%%%%%%%%%%%%%%%%%%%%%%%%%%%%%%%%%%%%%%%%%%
%
%  AUTHORS AND AFFILIATIONS
%
%%%%%%%%%%%%%%%%%%%%%%%%%%%%%%%%%%%%%%%%%%%%%%%

% Authors are individuals who have significantly contributed to the
% research and preparation of the article. Group authors are allowed, if
% each author in the group is separately identified in an appendix.)

% List authors by first name or initial followed by last name and
% separated by commas. Use \affil{} to number affiliations, and
% \thanks{} for author notes.
% Additional author notes should be indicated with \thanks{} (for
% example, for current addresses).

% Example: \authors{A. B. Author\affil{1}\thanks{Current address, Antartica}, B. C. Author\affil{2,3}, and D. E.
% Author\affil{3,4}\thanks{Also funded by Monsanto.}}

\authors{Charlotte C. Merchant\affil{1}, Milan Kl{\"o}wer\affil{1}, Bradley Stanley-Clamp\affil{2}, Maren H{\"o}ver\affil{1}, Simon L. L. Michel\affil{1}, Edward Groot\affil{1}, Hannah M. Christensen\affil{1}}

% \affiliation{1}{First Affiliation}
% \affiliation{2}{Second Affiliation}
% \affiliation{3}{Third Affiliation}
% \affiliation{4}{Fourth Affiliation}

\affiliation{1}{Department of Physics, University of Oxford, Oxford, UK}
\affiliation{2}{Department of Engineering, University of Oxford, Oxford, UK}
%(repeat as many times as is necessary)

% Corresponding author mailing address and e-mail address:

% (include name and email addresses of the corresponding author.  More
% than one corresponding author is allowed in this LaTeX file and for
% publication; but only one corresponding author is allowed in our
% editorial system.)

% Example: \correspondingauthor{First and Last Name}{email@address.edu}

\correspondingauthor{Charlotte Merchant}{charlotte.merchant@physics.ox.ac.uk}

%%%%%%%%%%%%%%%%%%%%%%%%%%%%%%%%%%%%%%%%%%%%%%%
% KEY POINTS
%%%%%%%%%%%%%%%%%%%%%%%%%%%%%%%%%%%%%%%%%%%%%%%
%  List up to three key points (at least one is required)
%  Key Points summarize the main points and conclusions of the article
%  Each must be 140 characters or fewer with no special characters or punctuation and must be complete sentences

% Example:
% \begin{keypoints}
% \item	List up to three key points (at least one is required)
% \item	Key Points summarize the main points and conclusions of the article
% \item	Each must be 140 characters or fewer with no special characters or punctuation and must be complete sentences
% \end{keypoints}

\begin{keypoints}
\item Most AI models reproduce present-day climate zones but misrepresent their redistribution in warmer climates.
\item AI models require explicit ocean-to-land coupling, as an architectural feature, to reproduce climate zone migration under warming.
\item We propose our climate-zone diagnostic as a validation test for AI climate projections.
\end{keypoints}

%%%%%%%%%%%%%%%%%%%%%%%%%%%%%%%%%%%%%%%%%%%%%%%
%
%  ABSTRACT and PLAIN LANGUAGE SUMMARY
%
% A good Abstract will begin with a short description of the problem
% being addressed, briefly describe the new data or analyses, then
% briefly states the main conclusion(s) and how they are supported and
% uncertainties.

% The Plain Language Summary should be written for a broad audience,
% including journalists and the science-interested public, that will not have 
% a background in your field.
%
% A Plain Language Summary is required in GRL, JGR: Planets, JGR: Biogeosciences,
% JGR: Oceans, G-Cubed, Reviews of Geophysics, and JAMES.
% see http://sharingscience.agu.org/creating-plain-language-summary/)
%
%%%%%%%%%%%%%%%%%%%%%%%%%%%%%%%%%%%%%%%%%%%%%%%

%% \begin{abstract} starts the second page

\begin{abstract}
Regional climate zones are expected to shift under global warming. 
Whether AI climate models have learned to generalize climate-zone distributions under warming in a physically meaningful way affects their suitability for climate projection. 
We address this question by applying a K{\"o}ppen-Geiger climate-zone decomposition to AIMIP Phase 1 models under prescribed $+4\,K$ SST forcing and comparing their responses to physics-based AMIP models. 
Using this diagnostic, we compare baseline classification skill, per-zone responses in temperature, precipitation, and near-surface specific humidity, and the spatial structure of departures from physics-based models. 
All AI models considered reproduce the 1979-2014 ERA5 climatology within the physics-based models' range, but only the hybrid physics-AI model NeuralGCM-HRD reorganizes zones in agreement with established thermodynamic and hydrological scaling relations. 
The remaining emulators have distinct failure modes traceable to their architectural treatment of land cells. 
A physically consistent climate-zone response is therefore necessary for AI models intended for climate projection.
\end{abstract}

\section*{Plain Language Summary}
The Earth has distinct regional climates, such as tropical, arid, and polar. 
From local long-term average temperature and precipitation, climate zones are defined following K{\"o}ppen-Geiger.
As the climate warms, these zones are expected to migrate, so a region that was once tropical may become arid.
Here, we test how well AI climate models capture current climate zones and the way those zones shift under warming.
The AI models we compare were only trained on data from 1979-2014, so whether they respond physically in an experiment with 4\textdegree{}\,K warmer ocean temperatures is unclear. 
We find that all AI models reproduce current climates zones as well as physics-based models, but under imposed warming, only a hybrid AI-physics model shifts its zones in agreement with physics-based models.
As such, we argue that new AI models should continue to be evaluated using similar methods as we use in this study.

%%%%%%%%%%%%%%%%%%%%%%%%%%%%%%%%%%%%%%%%%%%%%%%
%
%  BODY TEXT
%
%%%%%%%%%%%%%%%%%%%%%%%%%%%%%%%%%%%%%%%%%%%%%%%

%%% Suggested section heads:
% \section{Introduction}
%
% The main text should start with an introduction. Except for short
% manuscripts (such as comments and replies), the text should be divided
% into sections, each with its own heading.

% Headings should be sentence fragments and do not begin with a
% lowercase letter or number. Examples of good headings are:

% \section{Materials and Methods}
% Here is text on Materials and Methods.
%
% \subsection{A descriptive heading about methods}
% More about Methods.
%
% \section{Data} (Or section title might be a descriptive heading about data)
%
% \section{Results} (Or section title might be a descriptive heading about the
% results)
%
% \section{Conclusions}

\section{Introduction}
The K{\"o}ppen classification, devised in 1884 to map regional vegetation, partitions land into climate regimes based on thresholds of temperature, precipitation, and their seasonality \cite{koppen2011, Beck2018}. Under anthropogenic warming, climate zones are shifting and redistributing \cite{Chan2015}. K{\"o}ppen boundary migration integrates the joint thermal and hydrological response of the atmosphere over land into a single observable. The classification has been applied as a forced-response diagnostic to coupled general circulation models \cite{Lohmann1993,rubel2010,Belda2016,Beck2018} across successive generations of CMIP, the Coupled Model Intercomparison Project \cite{eyring2016, taylor2012}. The genesis of a new class of AI-based climate models, whose atmospheric physics is learned from reanalysis data (hereafter \emph{AI models}), offers an opportunity to apply K{\"o}ppen-based diagnostics as a test of out-of-distribution physical generalization, the ability to respond correctly to sea surface temperatures warmer than the training set. Do AI models overfit on present-day climate zones seen in training data, or have they learned to redistribute climate zones in a physically consistent way?

In lieu of integrating the discretized equations of motion found in traditional general circulation models (henceforth \emph{physics-based models}), AI models predict the atmosphere’s temporal evolution with neural networks, predominantly trained on reanalysis data. These AI models are built on different network architectures: from fully data-driven models that predict the entire atmospheric tendency \cite{Watt-Meyer2025}, to hybrid models that have a learned physics component but retain a dynamical core \cite{Kochkov2024, yuval2026}, to generative models that sample atmospheric states conditional on boundary forcing \cite{cBottle, hall2026}. The inductive biases embedded in these architectures differ accordingly, with strong physical priors in hybrid models and fewer priors in fully data-derived ones \cite{Karniadakis2021}. Several models were originally optimized for short-range weather prediction \cite{pathak2022, Bi2023, lam2023, lang2024}, but advances in rollout stability have made climate-length integrations tractable \cite{karlbauer, cresswell}. Once trained, an AI model can produce a multi-decadal simulation at a fraction of the computational cost of a physics-based model; AI models often take time steps of hours \cite{Bi2023, lam2023} compared to the seconds or minutes needed for numerical stability in physics-based models \cite{durran2010}. Whether efficient AI-based simulation of the observed climate translates into a physically plausible response to out-of-distribution forcing remains an open question.

The Artificial Intelligence Model Intercomparison Project (AIMIP Phase 1) extends the CMIP diagnostic framework to AI climate models \cite{henn2026}. AIMIP specifies a protocol similar to that of the (physics-based) Atmospheric Model Intercomparison Project (AMIP \cite{eyring2016}); participating models are forced by an ERA5-derived monthly sea surface temperature (SST) and sea-ice concentration dataset and integrated from 1978 through 2024 as a five-member inference ensemble \cite{henn2026}. Training is restricted to ERA5 historical observations from 1979-2014, which leaves 2015-2024 as an out-of-sample window within the historical record. Carbon dioxide and other radiative forcings are withheld as model inputs, so the prescribed SST forcing is the only warming signal available to the model. Two additional experiments apply uniform SST increases of $+2\,K$ and $+4\,K$ under an otherwise identical configuration. These warm-climate experiments are out-of-distribution by construction, so the simulated response is a test of extrapolation beyond the training distribution's support. Through common boundary forcings and evaluation periods, AIMIP enables explicit comparison across AI models, regardless of their network architecture or training strategies.

Previous evaluations of AI models, such as WeatherBench, have focused on the historical period and used numerical weather prediction baselines \cite{rasp2020, rasp2024}. Leading AI models reproduce mean climatologies \cite{Kochkov2024, Watt-Meyer2025, cresswell} and atmospheric variability such as tropical cyclone statistics \cite{lam2023, demaria2025aiwp} with comparable skill to physics-based models of similar horizontal resolution \cite{rasp2024, Kochkov2024, Watt-Meyer2025}. Such evaluations are necessary but limited in scope. Where the training data and forcing are from the historical period, in-sample skill conflates a learned physical response with historical memorization. The few studies that do consider the warm-climate behavior of AI models report two relevant signals. Both cBottle \cite{cBottle} and ACE2 \cite{Watt-Meyer2025} produce a weaker change in global-mean surface air temperature than expected given the imposed SST increase \cite{zhang2026}. Underestimating deviations from the training data is consistent with documented cold biases that emerge in AI models as warming proceeds historically \cite{rackow2024, landsberg2026, zhang2026}. As well, ACE2, NeuralGCM \cite{Kochkov2024}, and cBottle reproduce the wet-get-wetter, dry-get-drier response qualitatively but underestimate extreme precipitation scaling relative to Clausius-Clapeyron as shown in a GFDL-AM4, physics-based reference \cite{zhang2026}. K{\"o}ppen-zone decomposition localizes these signals, where we can trace the prescribed SST forcing through learned teleconnections and land-surface feedbacks into the climatic variables that define each zone. Thus, we investigate whether AI climate models have learned to generalize climate-zone redistributions under global warming in a physically consistent way.

In this study, we classify ERA5, AMIP, and AIMIP model climatologies by their dominant K{\"o}ppen classes and compare their responses under global warming experiments. Section \ref{methods} describes the datasets, models, and K{\"o}ppen classifications. Section 3 presents the geographic redistribution of zone boundaries, the class-mean response of physical variables (surface temperature, precipitation, and humidity), and distributional changes within zones in response to global warming relative to physics-based model responses. Implications for AI model development are discussed in Section 4.

\section{Data, Models, and Methods}
\label{methods}
\subsection{Data}
We compare data from three sources. ERA5 is our historical reference \cite{hersbach}. AMIP models provide a physics-based reference for the global warming experiments, where prescribed SSTs are warmed uniformly by $+2\,K$ and $+4\,K$ relative to a present-day baseline with no additional warming (+0\,K). AI models evaluated here, drawn from AIMIP, are trained on ERA5 over the historical period and forced with prescribed SSTs, so we can directly compare their warming response against the physics-based responses.

\subsubsection{ECMWF Reanalysis Version 5 (ERA5)}
ERA5 is the fifth-generation ECMWF global atmospheric reanalysis, produced by assimilating in-situ and remotely sensed observations from 1940 to present into ECMWF's weather model IFS \cite{hersbach}, and is used here in three respects. It is the training data for all AIMIP models; the source from which the prescribed monthly SST and sea-ice concentration forcing used by AIMIP during inference is derived \cite{henn2026}; and the present-day reference for each model’s $+0\,K$ K{\"o}ppen classification and per-zone density analysis (Figure \ref{fig01}; Figure S1). 

The relevant fields are 2-meter air temperature, surface precipitation rate, and 2-meter specific humidity (computed from 2-meter dewpoint where necessary). We express precipitation rate in mm month$^{-1}$ and follow CMIP6 conventions. We regrid ERA5 from its native 0.25\textdegree\, to a regular 1\textdegree\,longitude-latitude grid and restrict to cells that are at least 50\% land, using ERA5's fractional land-sea mask. All AI and physics-based model outputs are regridded to this same 1\textdegree\, grid prior to classification. 

\subsubsection{Atmospheric Model Intercomparison Project (AMIP)}
AMIP is the CMIP-coordinated protocol where atmosphere-only general circulation models are forced with prescribed observed monthly SSTs and sea-ice concentration \cite{eyring2016}.
Their amip-p4K and amip-m4K experiments are called $+4$\,K and $-4$\,K here.
We include any CMIP6 AMIP submission (listed in Table S1) that provides near-surface air temperature and surface precipitation for the relevant scenario over 1979-2014 and take the ensemble mean for each model.
The min-max range across AMIP models gives the spread of plausible physics-based responses to the same SST perturbation, and we evaluate the AIMIP responses against this range. Two AMIP models, CESM2 and GFDL-CM4, were selected as individual AMIP references because they have complete scenario coverage and have been analyzed in previous work \cite{zhang2026, landsberg2026}.

\subsection{Artificial Intelligence Model Intercomparison Project (AIMIP)}
AIMIP Phase 1 coordinated AMIP-style simulations of AI models exclusively trained on ERA5 from 1979-2014 \cite{henn2026}. The protocol reserves 2015-2024 as an out-of-sample test set and withholds CO$_2$ and other radiative forcings as inputs, so the prescribed SST, sea-ice concentration, and insolation seasonal cycle are the only time-varying forcings \cite{henn2026}.
We use the five AIMIP Phase 1 submissions \cite{henn2026}, enumerated below and summarized in Table S2.
All provide the minimum variable set for the K{\"o}ppen classification (2-meter air temperature and surface precipitation).
ArchesWeather and ArchesWeatherGen \cite{arches} are excluded as they did not provide surface precipitation at the time of analysis.
Each model contributes five ensemble members, each a single continuous simulation spanning October 1978 through December 2024 under the prescribed monthly SST and sea-ice forcing with monthly-mean output.

\subsubsection{Ai2 Climate Emulator Version 2 (ACE2.1-ERA5)}
ACE2.1-ERA5 is a variant of ACE2, a fully data-driven autoregressive deep-learning atmospheric emulator built on the Spherical Fourier Neural Operator (SFNO) \cite{Watt-Meyer2025, sfno}.
It integrates the global atmosphere forward on a 6-hour timestep with an architectural corrector that hard-conserves global dry-air mass and atmospheric moisture.
The model is deterministic, so ensemble members are generated by lagging the initial condition across five consecutive dates centered on 1 October 1978.
At inference, the prescribed SST forcing is applied by overwriting the prognostic surface temperature variable in majority-ocean cells, leaving surface temperature over land and sea-ice cells to evolve from the prognostic state \cite{henn2026}. Relative to ACE2, to comply with the AIMIP protocol, ACE2.1-ERA5 is trained without CO$_2$ as a forcing input, demotes near-surface variables to diagnostic outputs, conforms to the specified training period, and modifies the norm in the SFNO blocks \cite{henn2026}.

\subsubsection{Deep Learning Earth SYstem Model (DLESyM)}
DLESyM is a fully data-driven autoregressive coupled atmosphere-ocean model \cite{dlesym}.
The atmospheric component assessed in this study is a U-Net \cite{unet} of ConvNeXt \cite{liu2022} blocks and gated recurrent units with a 6-hour timestep and predicts the next state from the current and previous states.
Ensemble members follow the same initial-condition lagging scheme used for ACE2.1-ERA5.
For the AIMIP experiments, the prescribed SST forcing replaces the ocean component, which is zonally interpolated to infill land and sea-ice cells (to avoid masked cells in the model architecture), and the uniform perturbation is applied to the entire interpolated field \cite{henn2026}.
Consequently, DLESyM is the only AI model that imposes the SST perturbation directly on land and ice in our evaluation.
The checkpoint used from \citeA{cresswell} was trained from 1983 into 2016, which overextends the AIMIP holdout period, and included outgoing long-wave radiation from satellite observations, though, for the protocol, this was initialized with top net thermal radiation from ERA5 \cite{henn2026}.
DLESyM does not output near-surface humidity, so it is excluded from the Clausius-Clapeyron scaling analysis below.

\subsubsection{Neural General Circulation Model-[High Resolution Downscaling] (NeuralGCM-HRD)}
NeuralGCM is a hybrid physics-AI model that couples a differentiable, physics-based dynamical core to learned parametrizations trained on the residual against ERA5 \cite{Kochkov2024}.
NeuralGCM-HRD adds a learned decoder to downscale the 2.8\textdegree\, prognostic state to 1\textdegree\, surface and pressure-level outputs \cite{yuval2026}.
We use NeuralGCM-HRD for spatial consistency with other models. Ensemble members are generated by sampling the Gaussian Random Field input to the decoder heads.
Like ACE2.1-ERA5, NeuralGCM-HRD merges the prescribed ocean SST forcing with its learned land and sea-ice embeddings to form a complete surface field.
Because the SST perturbation is not applied directly to land or sea-ice cells, the land-warming response is generated by the model.
Surface precipitation is diagnosed as the time-mean column water vapor budget's residual, which combines the dynamical core's net precipitation with the surface head's predicted evaporation and is subject to a non-negativity constraint \cite{henn2026}.

\subsubsection{Climate in a Bottle (cBottle)}
cBottle is a conditional generative diffusion-based foundation model for the global atmosphere \cite{cBottle}.
Instead of propagating a prognostic state forward in time, cBottle samples atmospheric states from a learned conditional distribution, conditioned on prescribed SST and sea-ice fields.
It is a two-stage denoising diffusion model; cBottle-3d generates a coarse 1\textdegree\, global field, and cBottle-sr super-resolves it to $\sim$5km.
The AIMIP submission uses cBottle-3d on a HEALPix grid with 8 snapshots per day \cite{henn2026}.
SST fields are filled with a constant (290\,K) over land and sea-ice and unchanged during the experiments \cite{henn2026}, so cBottle receives no SST signal over land.
Sampling is regularized by three denoiser checkpoints spanning low, medium, and high diffusion noise levels.
Four ensemble members were generated by combining different checkpoint triples with diffusion latent representations that are correlated in time through a first-order autoregressive process with a 24-hour half-life.
The fifth ensemble member combined a checkpoint triple with a fully independent latent representation.
Only a single ensemble member is available for the warming experiments.

\subsubsection{Monthly Diffusion v0.9 (MD-1p5)}
MD-1p5 is an autoregressive latent-diffusion atmospheric model that advances a monthly-mean atmospheric state with a monthly timestep \cite{hall2026}.
The architecture is an encoder-predictor-decoder configuration using spectral convolution layers inspired by SFNOs.
A conditional variational autoencoder maps the prognostic state to a Gaussian latent posterior conditioned on the external forcings, a learned monthly seasonality embedding, and time-invariant orography.
The predictor is a conditional latent denoising diffusion probabilistic model that evolves the latent one month forward in time, and the decoder maps the result back to the monthly-mean prognostic state.
Over land and sea-ice cells, the SST input is assigned its climatological mean and held fixed across perturbation experiments, so no SST signal is applied to land.
The model is trained on ERA5 monthly means at 1.5\textdegree\, and regridded to 1\textdegree\, for the AIMIP submission.
Ensemble members are derived from different stochastic latent-diffusion samples, initialized from ERA5 in October 1978, and integrated through the end of 2024.

\subsection{Methods}
\subsubsection{K{\"o}ppen-Geiger Climate Classification}
The K{\"o}ppen-Geiger scheme partitions the land surface into discrete climate zones using thresholds of monthly mean near-surface temperature and precipitation.
Developed by \citeA{koppen1884} and refined by \citeA{geiger1954}, it assigns each land grid cell into one of five major classes, Tropical (A), Arid (B), Temperate (C), Cold (D), or Polar (E), based on the annual cycle of monthly mean temperatures and an aridity criterion.
Additional thresholds define a more granular set of 30 subclasses as introduced by \citeA{geiger1954}, but in this study, we work at the level of the five major classes.
The subclass partitions are more susceptible to small biases in precipitation seasonality, which may be present in AMIP, ERA5, and consequently the AI models.
We adopt the threshold table of \citeA{Beck2023} compiled from observations and constrained CMIP6 projections from 1901-2099 and apply it across ERA5, AMIP, and AIMIP model output at the $-4\,K$, $+0\,K$, $+2\,K$, and $+4\,K$ scenarios, classifying each year independently and assigning each cell its modal major class across 1979-2014, the period common to all three data sources. Per-model classification confidence plots are in Figure A1.

\begin{figure}
\noindent\includegraphics[width=\textwidth]{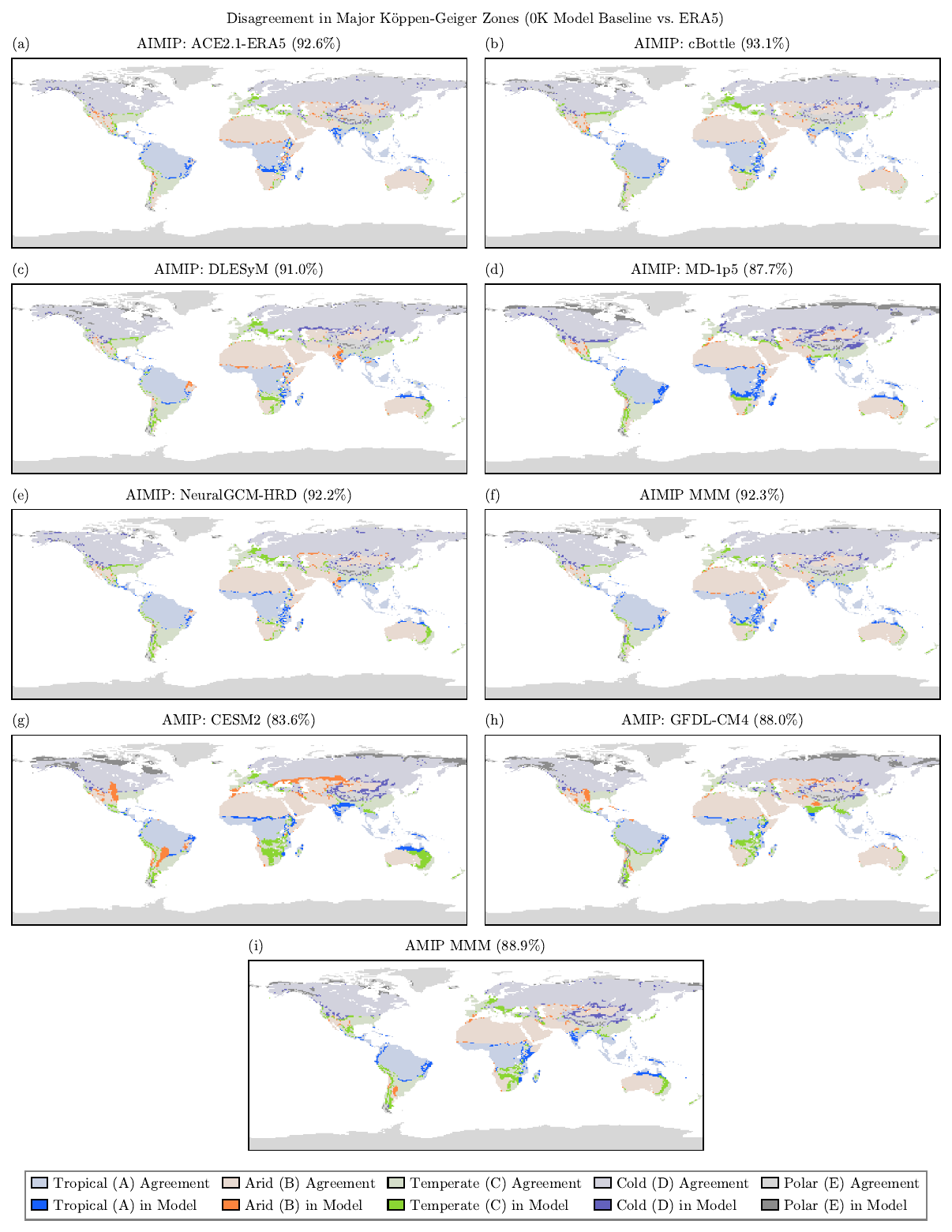}
\caption{\textbf{Differences in K{\"o}ppen-Geiger major climate zone classification between ERA5 and models' $\mathbf{+0\,K}$ baseline (1979-2014)}. Panels show the differences between ERA5's and the AIMIP and AMIP models' simulated baseline climate zones, computed from their respective ensemble mean. Subtitles report models' percent agreement with ERA5, computed as the cosine-latitude-weighted fraction of land assigned the same major zone. Across the AMIP ensemble, per-model agreement ranges from 79.0\% (NorCPM1) to 90.9\% (EC-Earth3-CC).}
\label{fig01}
\end{figure}

\section{Results}
In the multi-model mean, the five AI models reproduce the 1979-2014 ERA5 K{\"o}ppen-Geiger climatology over the training period better than the physics-based AMIP models (92.3\% versus  88.9\%; Figure \ref{fig01}), though MD-1p5 is a low outlier at 87.7\% agreement with ERA5. 
Disagreement is confined to threshold-sensitive boundaries, such as orographically complex regions and the Asian and African monsoon margins.
Because the AI models were trained on this period from ERA5, this agreement may reflect effective memorization of the training data rather than predictive skill, and the AIMIP and AMIP ensembles are not meaningfully separable on this metric.
Thus, the response to boundary forcing outside the training distribution is of greater diagnostic interest, and the remainder of this section investigates the uniform SST perturbation experiments.

A uniform $+4\,K$ SST perturbation is not a radiative forcing; the land only warms through the atmosphere's response to the imposed ocean warming, so a model does not need to reproduce the same land response as a coupled run whose radiative forcing increases SSTs by 4\,K.
Nonetheless, thermodynamic constraints apply.
Lower relative humidity over land steepens the change in lapse rate relative to the ocean, so depending on soil moisture, land generally warms more than ocean by a factor of $\sim$1.5 \cite{sutton2007}.
At high latitudes, lapse-rate and surface-albedo feedbacks amplify this warming further \cite{Pithan2014}.
Independent of circulation changes, a warmer atmosphere holds more water vapor, which reinforces the existing contrast between wet and dry regions \cite{held2006}; this wet-get-wetter tendency is attenuated over land, wherever moisture supply is rate-limited by upstream ocean evaporation \cite{byrne2015, byrne2016}.
Therefore, precipitation should intensify across all zones except Arid, most strongly in Cold and Polar zones \cite{held2006, seager2010}.
Where relative humidity is conserved, near-surface specific humidity should increase at the Clausius-Clapeyron rate ($\sim$7\%/K) \cite{held2006}, but where it declines, as over land, the increase is sub-Clausius-Clapeyron \cite{byrne2016, byrne2018}.
Translating this into the K{\"o}ppen-Geiger classification, we anticipate a poleward and upslope expansion of the warm and arid classes (A, B, C) and a contraction of cold, continental classes (D, E) \cite{Chan2015}.
The greatest fractional reorganization should occur in classes with the most land near a defining threshold, where the imposed warming reclassifies the largest area.

\begin{figure}
\noindent\includegraphics[width=\textwidth]{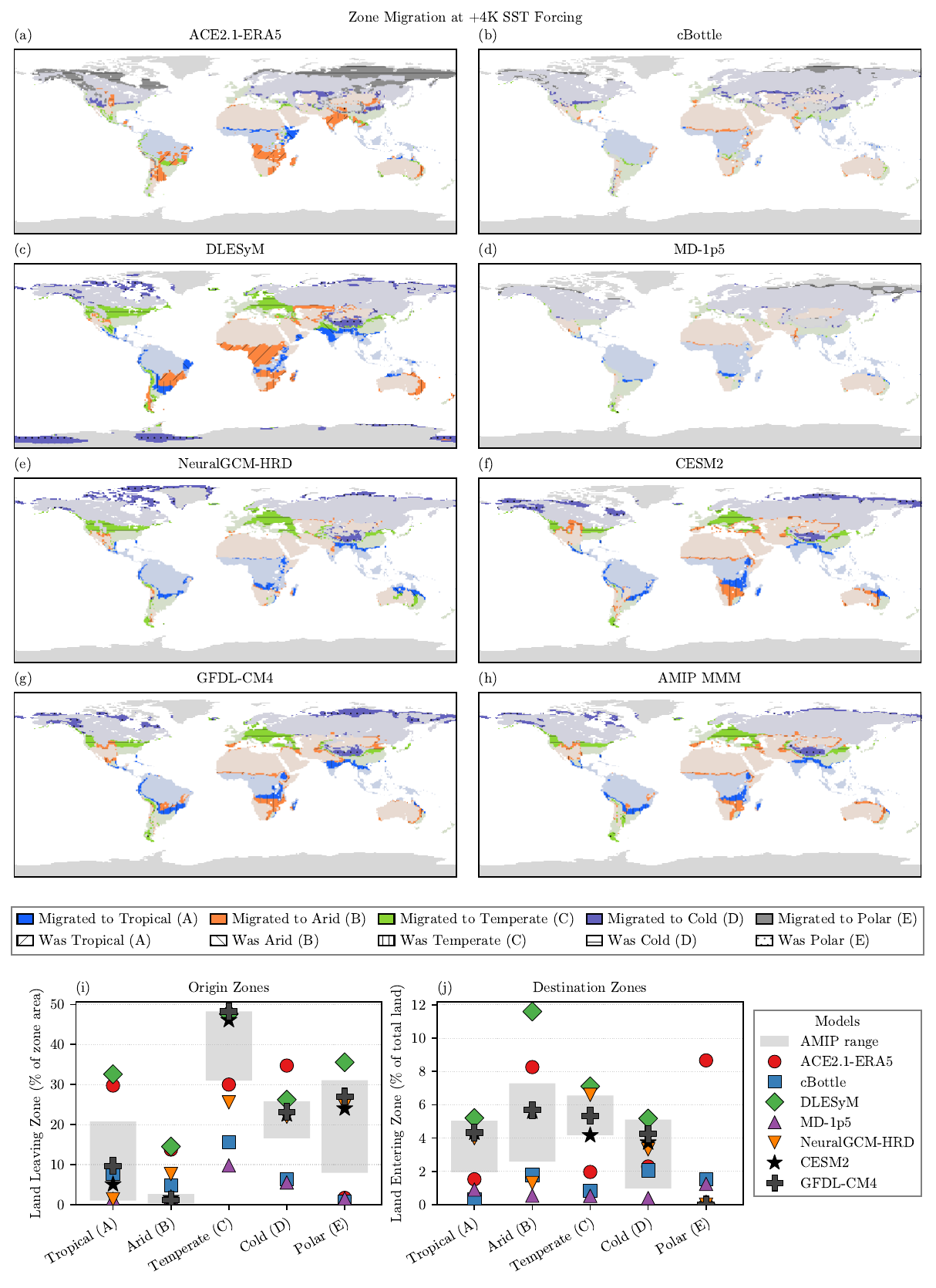}
\caption{\textbf{Climate zone migration at $\mathbf{+4\,K}$ SST forcing.} Land where K{\"o}ppen-Geiger major class differs between the $+0\,K$ baseline (1979-2014) and the $+4\,K$ SST warming experiment for the AIMIP models (a-e), representative AMIP models (f-g), and AMIP multi-model mean (MMM) (h). Migrated land is colored by destination climate zone with hatches indicating original zone.
Unchanged climate zones are muted.
To quantify the migration direction, Panels (i-j) show the per-zone land fraction leaving each origin zone (i) and the fraction of land entering each destination zone (j). Points mark AIMIP and named AMIP models against the grey AMIP ensemble min-max range.}
\label{fig02}
\end{figure}

The physics-based model ensemble realizes this expectation. Between 33 and 48\% of Temperate land and $\sim$20\% of Cold and Polar land migrates to a new major zone in the multi-model mean (Figure \ref{fig02}h).
Tropical and Arid redistribution is smaller because their present-day monthly-mean temperatures generally lie further from the relevant class boundaries than the magnitude of the applied perturbation.
The AI model responses divide into those that migrate climate zones in the expected direction in some regions (DLESyM and NeuralGCM-HRD), those that fail to migrate (cBottle and MD-1p5), and one whose migration opposes our physical expectations (ACE2.1-ERA5).

DLESyM reorganizes its zones in physically expected directions. 
Its arid expansion across the Sahel is consistent with a thermodynamic intensification of the wet-dry contrast at the subtropical dry belt's southern margin \cite{held2006}; its Temperate encroachment into eastern Europe and the northern contiguous United States reflects mid-latitude land warming that exceeds the prescribed SST change \cite{Joshi2008}, and the poleward retreat of the Cold class margins follows from high-latitude amplification \cite{Pithan2014, Beck2018} (Figure \ref{fig02}c). 
However, DLESyM is the only AI model where the SST perturbation is applied directly to land and sea-ice cells (Section \ref{methods}) \cite{dlesym, henn2026}, so part of its cryospheric response is prescribed rather than emergent.
NeuralGCM-HRD reorganizes in the same direction but at reduced amplitude. The model's weaker warming over Temperate land fails to displace it across the K{\"o}ppen-Geiger thresholds at AMIP-like rates (Figure \ref{fig02}i), and it misses the tropical-to-arid transition over southern Africa (Figure \ref{fig02}e).

cBottle and MD-1p5 show negligible migration across the Temperate, Cold, and Polar classes in Figure \ref{fig02}.
This absence is likely structural as both architectures hold land and sea-ice cells at a constant or climatological value during inference (Section \ref{methods}), so no internal mechanism propagates the SST signal onto land.
The lack of migration suggests absent regional forcing, not an absence of response.

\begin{figure}
\noindent\includegraphics[width=\textwidth]{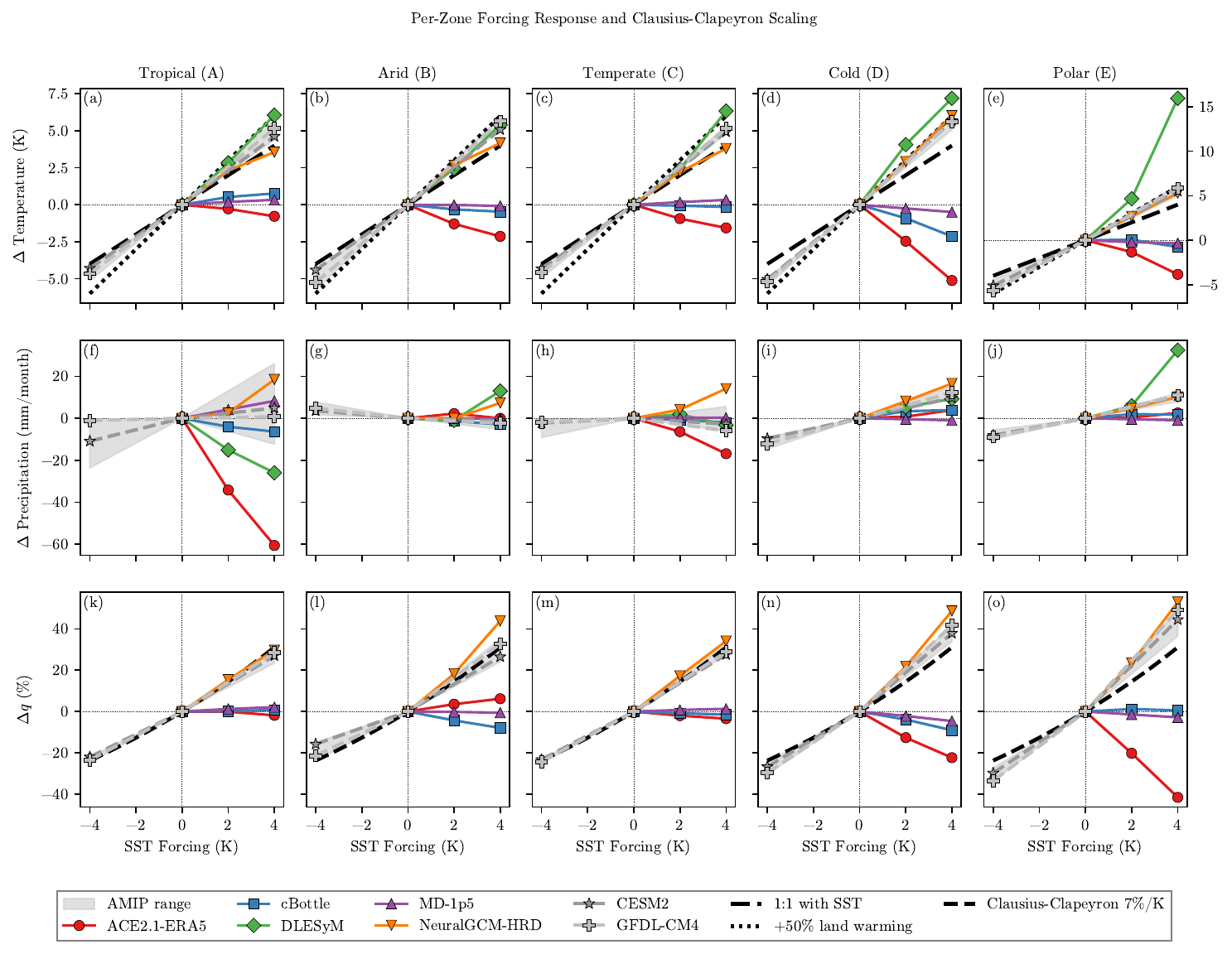}
\caption{\textbf{Response to uniform SST forcing per climate zone}.
Model responses in temperature (a-e), precipitation rate (f-j), and specific humidity (k-o), relative to the models' 0\,K baseline, versus prescribed SST forcing ($-4\,K$, $+0\,K$, $+2\,K$, and $+4\,K$) across the five climate zones (A-E). Colored lines denote individual AI models. The grey envelope spans the AMIP min-max range with two reference AMIP models (CESM2, GFDL-CM4) as dashed grey curves. Black reference lines mark 1:1 and $+50\%$ land warming with SST (a-e) and Clausius-Clapeyron 7\%/K humidity scaling (k-o).
Models without specific humidity are omitted from (k-o).}
\label{fig03}
\end{figure}

ACE2.1-ERA5’s redistribution is partially consistent with physical expectations and partially counter to them (Figure \ref{fig02}a).
While the conversion of mid-latitude Tropical cells to Arid is, in principle, consistent with subtropical drying or a poleward Hadley expansion \cite{held2006, Lu2007},
the model's underlying tropical response is a surface cooling and drying (Figure \ref{fig03}a).
The migration more likely reflects the emulator's erroneous tropical response than either mechanism.
The Polar margin's equatorward expansion into formerly Cold regions contradicts the physical expectation and the physics-based models, which agree on increased warming in polar regions (Figure \ref{fig02}).

The migration metric is discrete by construction.
A response may not register in climate zone reclassification if its amplitude falls below the K{\"o}ppen-Geiger-defined boundary.
To recover this sub-threshold structure, we analyze the temperature, precipitation and humidity response within each zone across the $-4\,K$, $+0\,K$, $+2\,K$ and $+4\,K$ SST experiments (Figure \ref{fig03}).
Physics-based models agree with our expectations across the $-4\,K$ to $+4\,K$ range of perturbations (Figure \ref{fig03}).

Only NeuralGCM-HRD reproduces all three expected thermodynamic responses.
The model's temperature, precipitation, and humidity responses lie within or near the physics-based model ensemble, and its specific humidity change approximates the 7\%/K Clausius-Clapeyron reference in Tropical and Temperate zones and exceeds it in Arid, Cold, and Polar at $+4\,K$.
DLESyM warms in every zone but over-warms the polar band by $\sim$10\,K at $+4\,K$ (250\% of the imposed ocean warming against the $\sim$150\% reference \cite{sutton2007}) and has a similarly inflated Polar precipitation increase.
Direct application of the SST perturbation to the cryosphere accounts for part of this response, but the magnitude indicates an overly strong surface-albedo feedback or an under-active negative lapse-rate feedback \cite{Pithan2014}. 
cBottle and MD-1p5 produce no discernible signal in any variable, zone, or forcing level.
ACE2.1-ERA5 responds with the wrong sign in several zones and displays Tropical surface cooling and drying at $+4\,K$. Its 50\% decrease in Polar near-surface specific-humidity is opposite in sign to and nearly twice the magnitude of the predicted Clausius-Clapeyron scaling, so the learned Cold-region moisture behavior runs counter to the temperature dependence of the saturation vapor pressure curve.

\begin{figure}
\centering
\noindent\includegraphics[width=\textwidth]{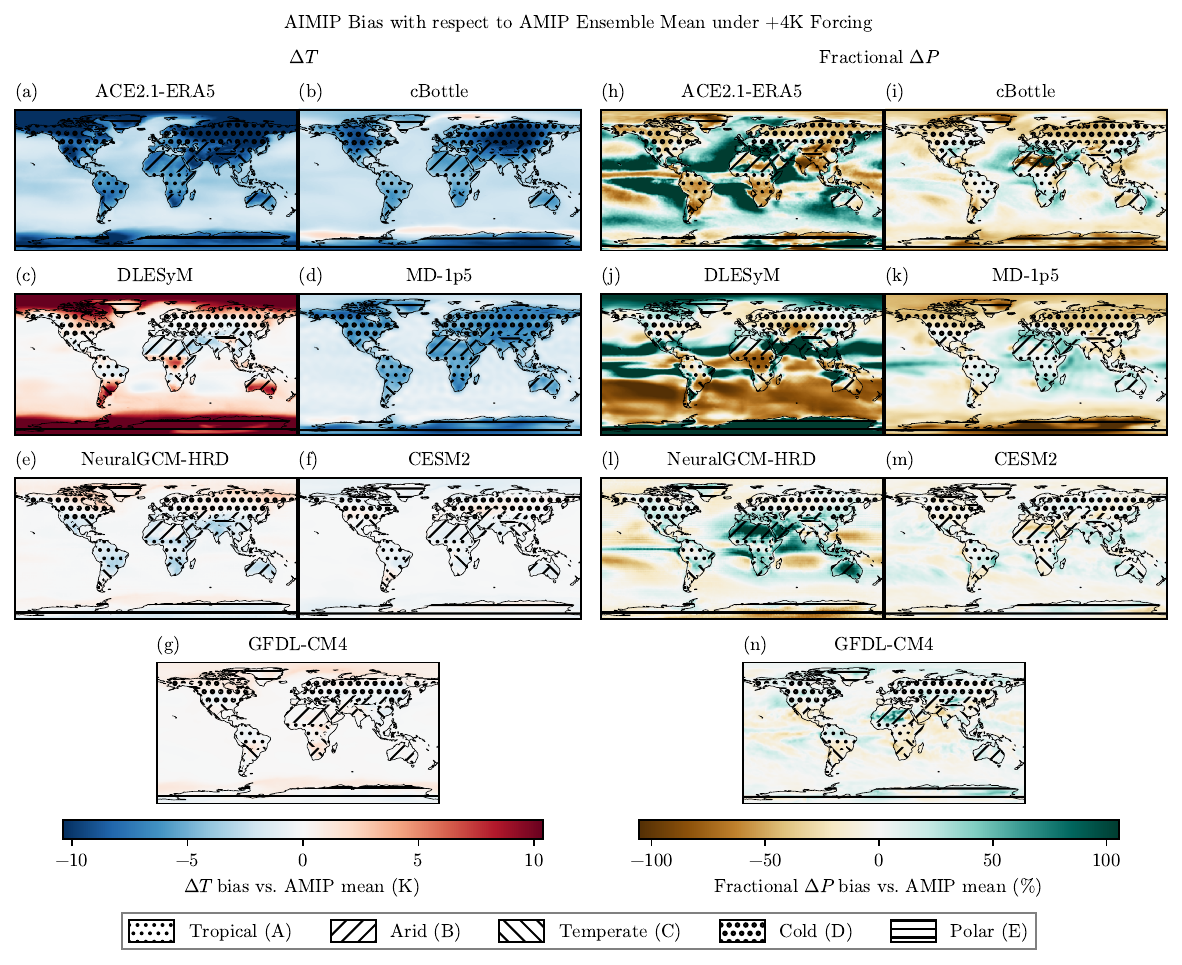}
\caption{\textbf{AI models' forcing response relative to physics-based models.} Per-model discrepancy in the $+4\,K$ response between each model (rows) and the AMIP multi-model mean for temperature (left, $\Delta T_{\text{AI}} - \Delta T_{\text{physics}}$, K) and fractional precipitation (right, \%). Each response is the ensemble-mean, time-mean difference between $+4\,K$ and the $+0\,K$ baseline experiments. White indicates agreement. For temperature, blue (red) indicates an under- (over-) response; for precipitation, brown (green) marks a drier (wetter) bias. Color limits are the 95th percentile of the absolute discrepancy across models, per variable. Each model's baseline K{\"o}ppen-Geiger climate zones are overlaid as hatching.}
\label{fig04}
\end{figure}

We now map the spatial structure of these discrepancies against the AMIP multi-model mean (Figure \ref{fig04}). 
Temperature discrepancies are small over the tropical and subtropical oceans, where prescribed SSTs tightly constrain the atmospheric response, and concentrate over land and toward both poles, where surface warming depends on feedbacks and land-atmosphere coupling.
Fidelity where SSTs are prescribed allows us to attribute divergence elsewhere to a failure of internally generated physics as opposed to structural bias.
Precipitation differences are less spatially organized but are generally elevated along the Intertropical Convergence Zone.
NeuralGCM-HRD is closest to the physics-based reference and only has localized departures.
ACE2.1-ERA5 exhibits a cold bias that intensifies toward both poles, a large dry bias over land, and a wet bias over the subtropical ocean.
cBottle presents a globally distributed cold and dry bias.
DLESyM shows the reverse in temperature, a near-global warm bias that intensifies poleward, alongside a hemispherically split precipitation bias, wet in the north and dry in the south, and a strong dry bias over Antarctica.
MD-1p5 is broadly cold over land and ocean, but its dry bias is confined to Cold and Polar zones.

\section{Conclusion and Discussion}
Most AI models in this study do not respond correctly to imposed ocean-surface warming (Figures \ref{fig03} and \ref{fig04}).
The true response is unknown, but it is unlikely far outside the AMIP uncertainty.
Under $+4\,K$ SST forcing, the AI models either respond too weakly compared to AMIP, or respond incorrectly, for example by excessively warming the polar regions or spuriously drying the surface air across many climate zones (Figure \ref{fig04}).
These responses conflict with physical understanding of the climate system and physics-based models (Figure \ref{fig03}). 
The AI models fit the training distribution well (Figure \ref{fig01}), but nearly all fail to generalize beyond it, which is consistent with \citeA{henn2026}.

Of the five AI models considered, the hybrid physics-AI model NeuralGCM-HRD is the only one to generalize K{\"o}ppen-Geiger climate zone distributions under $+4\,K$ SST forcing in a manner consistent with known approximate scaling relations of warming (e.g., Clausius-Clapeyron) and close to the physics-based model spread.
Presumably, NeuralGCM-HRD owes this physically realistic response to its explicit, learned coupling between SST and the land atmosphere, embedded within its physics-based dynamical core, which the other AI models lack.
cBottle and MD-1p5 hold land cells fixed, so the boundary perturbation does not propagate onto land.
ACE2.1-ERA5 lets land evolve prognostically but does not impose a constraint linking land temperature to the prescribed SST; its coupling is implicit in training-time correlations between land and ocean states that do not extrapolate to a perturbed boundary. 
DLESyM circumvents the question by applying the perturbation directly to its land and sea-ice cells, so part of its response is prescribed instead of learned. 
Consequently, how prescribed inputs propagate to prognostic variables is an architectural decision rather than an emergent property of training.
More broadly, an AI model's skill in reproducing present-day climate does not guarantee its ability to generalize to unseen climates.

By pairing per-zone response to SST forcing with spatial comparison against the physics-based AMIP ensemble, our analysis exposes shortcomings in AI models' ability to extrapolate to warmer climates.
We therefore suggest our diagnostics are a prerequisite, not a supplement, for evaluating AI models intended for climate projections.

This investigation has several limitations arising from the design of the AIMIP protocol, which withholds radiative forcing and only defines SST forcing experiments.
Global warming is a coupled dynamical-thermodynamical response to increases in greenhouse gas concentrations, aerosols, and land-use change, and their implicit feedbacks.
In that sense, an AI model that performs well in the AIMIP experiments may not respond in a physically coherent way to other warming scenarios, including radiative and regional forcings. 
As well, ensemble sizes under the perturbation experiments are small, particularly for cBottle, and the physics-based AMIP reference is itself a set of models with their structural biases.
In future phases of AI model intercomparisons, our diagnostics should be applied to fully coupled models, as in \citeA{antonio2026}, under explicit radiative forcing to test how the response to imposed CO$_2$ propagates through the coupled ocean-atmosphere-land system and how it modulates the system's internal variability.
Until then, the models tested here must demonstrate physical generalization before they can be used for reliable climate projection and, ultimately, to inform climate policy.

\section*{Open Research Section}
Code to reproduce figures is available at \url{https://github.com/c-merchant/aimip1-generalisability} and will be converted to a DOI upon acceptance.
AIMIP Phase 1 \cite{henn2026} data is accessible via DKRZ and can be downloaded via instructions at  \url{https://github.com/ai2cm/AIMIP}.
ERA5 reanalysis \cite{hersbach} and AMIP data \cite{eyring2016} can be accessed through the Copernicus Climate Data Store (\url{https://cds.climate.copernicus.eu/}) and the ESGF CMIP6 archive (\url{https://esgf-ui.ceda.ac.uk/cog/search/cmip6-ceda/}) respectively. 

\section*{Conflict of Interest declaration}
The authors declare there are no conflicts of interest for this manuscript.

\section*{Author Contributions}
CM - Methodology, Formal analysis, Investigation, Data curation, Visualization, Writing - Original Draft;
MK - Conceptualization, Writing - Review \& Editing, Supervision;
BSC - Software, Writing - Review \& Editing;
MH - Validation, Writing - Original Draft, Writing - Review \& Editing;
SM - Data curation, Writing - Review \& Editing;
EG - Writing - Review \& Editing;
HC - Conceptualization, Resources, Writing - Review \& Editing, Supervision.

CM developed the methodology, carried out the formal analysis and investigation, curated the data, produced the figures and wrote the original draft. BSC developed the clustering module and associated tests. MH validated the analysis implementation and contributed to writing the original draft. SM contributed to data curation. MK and HC conceived the study and supervised the work. MK supervised the analysis throughout, and HC shaped the AIMIP protocol discussions and secured early access to model output. All authors reviewed and edited the manuscript.
\acknowledgments
This work was conducted as part of the Intelligent Earth CDT supported by funding from the UK Research and Innovation Council (UKRI) grant number EP/Y030907/1.
MK acknowledges funding from the Natural Environment Research Council under grant number UKRI191. MH acknowledges funding by the Rhodes Trust.

HMC and SLLM acknowledge funding from the EERIE project (Grant Agreement No 101081383) funded by the European Union: University of Oxford's contribution to EERIE is funded by UK Research and Innovation (UKRI) under the UK government’s Horizon Europe funding guarantee (grant number 10049639). Views and opinions expressed are however those of the author(s) only and do not necessarily reflect those of the European Union or the European Climate Infrastructure and Environment Executive Agency (CINEA). Neither the European Union nor the granting authority can be held responsible for them. HMC and EG were supported by the Leverhulme Trust Research Project Grant `Exposing the nature of model error in weather and climate models'. HMC was further supported by a Leverhulme Trust Research Leadership Award  `Seamless Uncertainty Quantification for Earth System prediction' (SUQCES). 

For the purpose of Open Access, the author has applied a CC BY public copyright licence to any Author Accepted Manuscript version arising from this submission.
%%%%%%%%%%%%%%%%%%%%%%%%%%%%%%%%%%%%%%%%%%%%%%%
% REFERENCES and BIBLIOGRAPHY
%
\bibliography{agusample}

@article{geiger1954,
  author  = {Geiger, R. and Pohl, W.},
  title   = {Eine neue Wandkarte der Klimagebiete der Erde nach {W. K{\"o}ppens} Klassifikation},
  journal = {Erdkunde},
  year    = {1954},
  volume  = {8},
  number  = {1},
  pages   = {58--61},
  doi     = {10.3112/erdkunde.1954.01.04},
  url     = {https://doi.org/10.3112/erdkunde.1954.01.04}
}

@article{koppen1884,
  author  = {K{\"o}ppen, W.},
  title   = {Die W{\"a}rmezonen der Erde, nach der Dauer der heissen, gem{\"a}ssigten und kalten Zeit und nach der Wirkung der W{\"a}rme auf die organische Welt betrachtet},
  journal = {Meteorologische Zeitschrift},
  year    = {1884},
  volume  = {1},
  pages   = {215--226},
  url     = {https://koeppen-geiger.vu-wien.ac.at/pdf/Koppen_1884.pdf}
}

@article{byrne2018,
author = {Michael P. Byrne  and Paul A. O’Gorman },
title = {Trends in continental temperature and humidity directly linked to ocean warming},
journal = {Proceedings of the National Academy of Sciences},
volume = {115},
number = {19},
pages = {4863-4868},
year = {2018},
doi = {10.1073/pnas.1722312115},
URL = {https://www.pnas.org/doi/abs/10.1073/pnas.1722312115},
eprint = {https://www.pnas.org/doi/pdf/10.1073/pnas.1722312115}}

@article {byrne2016,
      author = "Michael P. Byrne and Paul A. O’Gorman",
      title = "Understanding Decreases in Land Relative Humidity with Global Warming: Conceptual Model and GCM Simulations",
      journal = "Journal of Climate",
      year = "2016",
      publisher = "American Meteorological Society",
      address = "Boston MA, USA",
      volume = "29",
      number = "24",
      doi = "10.1175/JCLI-D-16-0351.1",
      pages=      "9045 - 9061",
      url = "https://journals.ametsoc.org/view/journals/clim/29/24/jcli-d-16-0351.1.xml"
}

@article{Bi2023,
  author  = {Bi, Kaifeng and Xie, Lingxi and Zhang, Hengheng and Chen, Xin and Gu, Xiaotao and Tian, Qi},
  title   = {Accurate medium-range global weather forecasting with 3D neural networks},
  journal = {Nature},
  year    = {2023},
  volume  = {619},
  number  = {7970},
  pages   = {533--538},
  month   = jul,
  issn    = {1476-4687},
  doi     = {10.1038/s41586-023-06185-3},
  url     = {https://doi.org/10.1038/s41586-023-06185-3}
}

@article{lam2023,
  author    = {Lam, Remi and Sanchez-Gonzalez, Alvaro and Willson, Matthew and Wirnsberger, Peter and Fortunato, Meire and Alet, Ferran and Ravuri, Suman and Ewalds, Timo and Eaton-Rosen, Zach and Hu, Weihua and Merose, Alexander and Hoyer, Stephan and Holland, George and Vinyals, Oriol and Stott, Jacklynn and Pritzel, Alexander and Mohamed, Shakir and Battaglia, Peter},
  title     = {Learning skillful medium-range global weather forecasting},
  journal   = {Science},
  volume    = {382},
  number    = {6677},
  pages     = {1416--1421},
  year      = {2023},
  doi       = {10.1126/science.adi2336},
  url       = {https://www.science.org/doi/abs/10.1126/science.adi2336},
  eprint    = {https://www.science.org/doi/pdf/10.1126/science.adi2336}
}

@book{durran2010,
language = {eng},
lccn = {2010934663},
publisher = {Springer},
series = {Texts in applied mathematics, 32},
title = {Numerical methods for fluid dynamics : with applications to geophysics },
url = {http://www.springer.com/gb/},
year = {2010},
author = {Durran, Dale R.},
address = {New York},
booktitle = {Numerical methods for fluid dynamics : with applications to geophysics},
edition = {2nd ed.},
isbn = {9781441964113},
}

@article {seager2010,
      author = "Richard Seager and Naomi Naik and Gabriel A. Vecchi",
      title = "Thermodynamic and Dynamic Mechanisms for Large-Scale Changes in the Hydrological Cycle in Response to Global Warming",
      journal = "Journal of Climate",
      year = "2010",
      publisher = "American Meteorological Society",
      address = "Boston MA, USA",
      volume = "23",
      number = "17",
      doi = "10.1175/2010JCLI3655.1",
      pages=      "4651 - 4668",
      url = "https://journals.ametsoc.org/view/journals/clim/23/17/2010jcli3655.1.xml"
}

@misc{lang2024,
      title={AIFS -- ECMWF's data-driven forecasting system}, 
      author={Simon Lang and Mihai Alexe and Matthew Chantry and Jesper Dramsch and Florian Pinault and Baudouin Raoult and Mariana C. A. Clare and Christian Lessig and Michael Maier-Gerber and Linus Magnusson and Zied Ben Bouallègue and Ana Prieto Nemesio and Peter D. Dueben and Andrew Brown and Florian Pappenberger and Florence Rabier},
      year={2024},
      eprint={2406.01465},
      archivePrefix={arXiv},
      primaryClass={physics.ao-ph},
      url={https://arxiv.org/abs/2406.01465}, 
}

@misc{pathak2022,
      title={FourCastNet: A Global Data-driven High-resolution Weather Model using Adaptive Fourier Neural Operators}, 
      author={Jaideep Pathak and Shashank Subramanian and Peter Harrington and Sanjeev Raja and Ashesh Chattopadhyay and Morteza Mardani and Thorsten Kurth and David Hall and Zongyi Li and Kamyar Azizzadenesheli and Pedram Hassanzadeh and Karthik Kashinath and Animashree Anandkumar},
      year={2022},
      eprint={2202.11214},
      archivePrefix={arXiv},
      primaryClass={physics.ao-ph},
      url={https://arxiv.org/abs/2202.11214}, 
}

@article{demaria2025aiwp,
  author    = {DeMaria, Mark and Franklin, James L. and Chirokova, Galina and Radford, Jacob and DeMaria, Robert and Musgrave, Kate D. and Ebert-Uphoff, Imme},
  title     = {An Operations-Based Evaluation of Tropical Cyclone Track and Intensity Forecasts from Artificial Intelligence Weather Prediction Models},
  journal   = {Artificial Intelligence for the Earth Systems},
  year      = {2025},
  publisher = {American Meteorological Society},
  address   = {Boston, MA, USA},
  volume    = {4},
  number    = {4},
  pages     = {240085},
  doi       = {10.1175/AIES-D-24-0085.1},
  url       = {https://journals.ametsoc.org/view/journals/aies/4/4/AIES-D-24-0085.1.xml}
}

@article{karlbauer,
author = {Karlbauer, Matthias and Cresswell-Clay, Nathaniel and Durran, Dale R. and Moreno, Raul A. and Kurth, Thorsten and Bonev, Boris and Brenowitz, Noah and Butz, Martin V.},
title = {Advancing Parsimonious Deep Learning Weather Prediction Using the HEALPix Mesh},
journal = {Journal of Advances in Modeling Earth Systems},
volume = {16},
number = {8},
pages = {e2023MS004021},
doi = {https://doi.org/10.1029/2023MS004021},
url = {https://agupubs.onlinelibrary.wiley.com/doi/abs/10.1029/2023MS004021},
eprint = {https://agupubs.onlinelibrary.wiley.com/doi/pdf/10.1029/2023MS004021},
note = {e2023MS004021 2023MS004021},
year = {2024}
}

@misc{antonio2026,
      title={ACE2-NEMO: Coupling an ML atmospheric emulator to a full-depth dynamical ocean model}, 
      author={Bobby Antonio and Kristian Strommen and Pablo Ortega and Hannah M. Christensen},
      year={2026},
      eprint={2603.28704},
      archivePrefix={arXiv},
      primaryClass={physics.ao-ph},
      url={https://arxiv.org/abs/2603.28704}, 
}

@article{rasp2020,
author = {Rasp, Stephan and Dueben, Peter D. and Scher, Sebastian and Weyn, Jonathan A. and Mouatadid, Soukayna and Thuerey, Nils},
title = {WeatherBench: A Benchmark Data Set for Data-Driven Weather Forecasting},
journal = {Journal of Advances in Modeling Earth Systems},
volume = {12},
number = {11},
pages = {e2020MS002203},
doi = {https://doi.org/10.1029/2020MS002203},
url = {https://agupubs.onlinelibrary.wiley.com/doi/abs/10.1029/2020MS002203},
eprint = {https://agupubs.onlinelibrary.wiley.com/doi/pdf/10.1029/2020MS002203},
note = {e2020MS002203 10.1029/2020MS002203},
year = {2020}
}

@misc{rasp2024,
      title={WeatherBench 2: A benchmark for the next generation of data-driven global weather models}, 
      author={Stephan Rasp and Stephan Hoyer and Alexander Merose and Ian Langmore and Peter Battaglia and Tyler Russel and Alvaro Sanchez-Gonzalez and Vivian Yang and Rob Carver and Shreya Agrawal and Matthew Chantry and Zied Ben Bouallegue and Peter Dueben and Carla Bromberg and Jared Sisk and Luke Barrington and Aaron Bell and Fei Sha},
      year={2024},
      eprint={2308.15560},
      archivePrefix={arXiv},
      primaryClass={physics.ao-ph},
      url={https://arxiv.org/abs/2308.15560}, 
}

@misc{hall2026,
      title={Monthly Diffusion v0.9: A Latent Diffusion Model for the First AI-MIP}, 
      author={Kyle J. C. Hall and Maria J. Molina},
      year={2026},
      eprint={2604.13481},
      archivePrefix={arXiv},
      primaryClass={cs.LG},
      url={https://arxiv.org/abs/2604.13481}, 
}

@article{Lohmann1993,
  author  = {Lohmann, U. and Sausen, R. and Bengtsson, L. and Cubasch, U. and Perlwitz, J. and Roeckner, E.},
  title   = {The {K\"o}ppen climate classification as a diagnostic tool for general circulation models},
  journal = {Climate Research},
  year    = {1993},
  volume  = {3},
  pages   = {177--193},
  issn    = {0936-577X}
}

@Article{rubel2010,
author = "Rubel, Franz and Kottek, Markus",
journal = "Meteorologische Zeitschrift",
month = 04,
year = 2010,
title = "Observed and projected climate shifts 1901-2100 depicted by world maps of the K{\"o}ppen-Geiger climate classification",
number = "2",
volume = "19",
pages = {135-141},
url = "http://dx.doi.org/10.1127/0941-2948/2010/0430",
doi = "10.1127/0941-2948/2010/0430",
publisher = "Schweizerbart Science Publishers",
address = "Stuttgart, Germany"
}

@article{held2006,
      author = "Isaac M. Held and Brian J. Soden",
      title = "Robust Responses of the Hydrological Cycle to Global Warming",
      journal = "Journal of Climate",
      year = "2006",
      publisher = "American Meteorological Society",
      address = "Boston MA, USA",
      volume = "19",
      number = "21",
      doi = "10.1175/JCLI3990.1",
      pages=      "5686 - 5699",
      url = "https://journals.ametsoc.org/view/journals/clim/19/21/jcli3990.1.xml"
}

@article{Lu2007,
author = {Lu, Jian and Vecchi, Gabriel A. and Reichler, Thomas},
title = {Expansion of the Hadley cell under global warming},
journal = {Geophysical Research Letters},
volume = {34},
number = {6},
pages = {},
doi = {https://doi.org/10.1029/2006GL028443},
url = {https://agupubs.onlinelibrary.wiley.com/doi/abs/10.1029/2006GL028443},
eprint = {https://agupubs.onlinelibrary.wiley.com/doi/pdf/10.1029/2006GL028443},
year = {2007}
}

@article{sutton2007,
author = {Sutton, Rowan T. and Dong, Buwen and Gregory, Jonathan M.},
title = {Land/sea warming ratio in response to climate change: IPCC AR4 model results and comparison with observations},
journal = {Geophysical Research Letters},
volume = {34},
number = {2},
pages = {},
doi = {https://doi.org/10.1029/2006GL028164},
url = {https://agupubs.onlinelibrary.wiley.com/doi/abs/10.1029/2006GL028164},
eprint = {https://agupubs.onlinelibrary.wiley.com/doi/pdf/10.1029/2006GL028164},
year = {2007}
}

@Article{eyring2016,
AUTHOR = {Eyring, V. and Bony, S. and Meehl, G. A. and Senior, C. A. and Stevens, B. and Stouffer, R. J. and Taylor, K. E.},
TITLE = {Overview of the Coupled Model Intercomparison Project Phase 6 (CMIP6)
experimental design and organization},
JOURNAL = {Geoscientific Model Development},
VOLUME = {9},
YEAR = {2016},
NUMBER = {5},
PAGES = {1937--1958},
URL = {https://gmd.copernicus.org/articles/9/1937/2016/},
DOI = {10.5194/gmd-9-1937-2016}
}

@article{hersbach,
author = {Hersbach, Hans and Bell, Bill and Berrisford, Paul and Hirahara, Shoji and Horányi, András and Muñoz-Sabater, Joaquín and Nicolas, Julien and Peubey, Carole and Radu, Raluca and Schepers, Dinand and Simmons, Adrian and Soci, Cornel and Abdalla, Saleh and Abellan, Xavier and Balsamo, Gianpaolo and Bechtold, Peter and Biavati, Gionata and Bidlot, Jean and Bonavita, Massimo and De Chiara, Giovanna and Dahlgren, Per and Dee, Dick and Diamantakis, Michail and Dragani, Rossana and Flemming, Johannes and Forbes, Richard and Fuentes, Manuel and Geer, Alan and Haimberger, Leo and Healy, Sean and Hogan, Robin J. and Hólm, Elías and Janisková, Marta and Keeley, Sarah and Laloyaux, Patrick and Lopez, Philippe and Lupu, Cristina and Radnoti, Gabor and de Rosnay, Patricia and Rozum, Iryna and Vamborg, Freja and Villaume, Sebastien and Thépaut, Jean-Noël},
title = {The ERA5 global reanalysis},
journal = {Quarterly Journal of the Royal Meteorological Society},
volume = {146},
number = {730},
pages = {1999-2049},
doi = {https://doi.org/10.1002/qj.3803},
url = {https://rmets.onlinelibrary.wiley.com/doi/abs/10.1002/qj.3803},
eprint = {https://rmets.onlinelibrary.wiley.com/doi/pdf/10.1002/qj.3803},
year = {2020}
}

@article{Joshi2008,
  author   = {Joshi, Manoj M. and Gregory, Jonathan M. and Webb, Mark J. and Sexton, David M. H. and Johns, Tim C.},
  title    = {Mechanisms for the land/sea warming contrast exhibited by simulations of climate change},
  journal  = {Climate Dynamics},
  year     = {2008},
  month    = apr,
  volume   = {30},
  number   = {5},
  pages    = {455--465},
  issn     = {1432-0894},
  doi      = {10.1007/s00382-007-0306-1},
  url      = {https://doi.org/10.1007/s00382-007-0306-1},
}

@article{Pithan2014,
  author    = {Pithan, Felix and Mauritsen, Thorsten},
  title     = {Arctic amplification dominated by temperature feedbacks in contemporary climate models},
  journal   = {Nature Geoscience},
  year      = {2014},
  month     = mar,
  volume    = {7},
  number    = {3},
  pages     = {181--184},
  issn      = {1752-0908},
  doi       = {10.1038/ngeo2071},
  url       = {https://doi.org/10.1038/ngeo2071}
}

@article {byrne2015,
      author = "Michael P. Byrne and Paul A. O’Gorman",
      title = "The Response of Precipitation Minus Evapotranspiration to Climate Warming: Why the “Wet-Get-Wetter, Dry-Get-Drier” Scaling Does Not Hold over Land",
      journal = "Journal of Climate",
      year = "2015",
      publisher = "American Meteorological Society",
      address = "Boston MA, USA",
      volume = "28",
      number = "20",
      doi = "10.1175/JCLI-D-15-0369.1",
      pages=      "8078 - 8092",
      url = "https://journals.ametsoc.org/view/journals/clim/28/20/jcli-d-15-0369.1.xml"
}

@article{Beck2023,
  author    = {Beck, Hylke E. and McVicar, Tim R. and Vergopolan, Noemi and Berg, Alexis and Lutsko, Nicholas J. and Dufour, Ambroise and Zeng, Zhenzhong and Jiang, Xin and van Dijk, Albert I. J. M. and Miralles, Diego G.},
  title     = {High-resolution (1 km) {K\"o}ppen-Geiger maps for 1901--2099 based on constrained {CMIP6} projections},
  journal   = {Scientific Data},
  year      = {2023},
  month     = oct,
  volume    = {10},
  number    = {1},
  pages     = {724},
  issn      = {2052-4463},
  doi       = {10.1038/s41597-023-02549-6},
  url       = {https://doi.org/10.1038/s41597-023-02549-6}
}

@article{arches,
author = {Guillaume Couairon  and Renu Singh  and Anastase Charantonis  and Christian Lessig  and Claire Monteleoni },
title = {ArchesWeatherGen: Skillful and compute-efficient probabilistic weather forecasting with machine learning},
journal = {Science Advances},
volume = {12},
number = {17},
pages = {eadx2372},
year = {2026},
doi = {10.1126/sciadv.adx2372},
URL = {https://www.science.org/doi/abs/10.1126/sciadv.adx2372},
eprint = {https://www.science.org/doi/pdf/10.1126/sciadv.adx2372}}

@article{cresswell,
author = {Cresswell-Clay, Nathaniel and Liu, Bowen and Durran, Dale R. and Liu, Zihui and Espinosa, Zachary I. and Moreno, Raul A. and Karlbauer, Matthias},
title = {A Deep Learning Earth System Model for Efficient Simulation of the Observed Climate},
journal = {AGU Advances},
volume = {6},
number = {4},
pages = {e2025AV001706},
doi = {https://doi.org/10.1029/2025AV001706},
url = {https://agupubs.onlinelibrary.wiley.com/doi/abs/10.1029/2025AV001706},
eprint = {https://agupubs.onlinelibrary.wiley.com/doi/pdf/10.1029/2025AV001706},
note = {e2025AV001706 2025AV001706},
year = {2025}
}

@article{Belda2016,
  author  = {Belda, M. and Holtanov{\'a}, E. and Kalvov{\'a}, J. and Halenka, T.},
  title   = {Global warming-induced changes in climate zones based on {CMIP5} projections},
  journal = {Climate Research},
  year    = {2016},
  volume  = {71},
  pages   = {17--31},
  doi     = {10.3354/cr01418},
  url     = {https://doi.org/10.3354/cr01418}
}

@misc{rackow2024,
      title={Robustness of AI-based weather forecasts in a changing climate}, 
      author={Thomas Rackow and Nikolay Koldunov and Christian Lessig and Irina Sandu and Mihai Alexe and Matthew Chantry and Mariana Clare and Jesper Dramsch and Florian Pappenberger and Xabier Pedruzo-Bagazgoitia and Steffen Tietsche and Thomas Jung},
      year={2024},
      eprint={2409.18529},
      archivePrefix={arXiv},
      primaryClass={physics.ao-ph},
      url={https://arxiv.org/abs/2409.18529}, 
}

@misc{henn2026,
      title={AIMIP Phase 1: systematic evaluations of AI weather and climate models}, 
      author={Brian Henn and Christopher S. Bretherton and Nikolay Kodunov and Christian Lessig and Maria J. Molina and Troy Arcomano and Oliver Watt-Meyer and Guillaume Couairon and Renu Singh and Robert Brunstein and Yana Hasson and Antonia Jost and Noah Brenowitz and Peter Manshausen and Nathaniel Cresswell-Clay and Dale Durran and Kyle Joseph Chen Hall and Janni Yuval and Dmitrii Kochkov and Stephan Hoyer and Ignacio Lopez-Gomez},
      year={2026},
      eprint={2605.06944},
      archivePrefix={arXiv},
      primaryClass={physics.ao-ph},
      url={https://arxiv.org/abs/2605.06944}, 
}

@article{yuval2026,
author = {Janni Yuval  and Ian Langmore  and Dmitrii Kochkov  and Stephan Hoyer },
title = {Neural general circulation models for modeling precipitation},
journal = {Science Advances},
volume = {12},
number = {2},
pages = {eadv6891},
year = {2026},
doi = {10.1126/sciadv.adv6891},
URL = {https://www.science.org/doi/abs/10.1126/sciadv.adv6891},
eprint = {https://www.science.org/doi/pdf/10.1126/sciadv.adv6891}}

@article{landsberg2026,
author = {Landsberg, Jacob B. and Barnes, Elizabeth A.},
title = {Forecasting the Future With Yesterday's Climate: Temperature Bias in AI Weather and Climate Models},
journal = {Geophysical Research Letters},
volume = {53},
number = {6},
pages = {e2025GL119740},
doi = {https://doi.org/10.1029/2025GL119740},
url = {https://agupubs.onlinelibrary.wiley.com/doi/abs/10.1029/2025GL119740},
eprint = {https://agupubs.onlinelibrary.wiley.com/doi/pdf/10.1029/2025GL119740},
note = {e2025GL119740 2025GL119740},
year = {2026}
}

@misc{zhang2026,
      title={The Equilibrium Response of Atmospheric Machine-Learning Models to Uniform Sea Surface Temperature Warming}, 
      author={Bosong Zhang and Timothy M. Merlis},
      year={2026},
      eprint={2510.02415},
      archivePrefix={arXiv},
      primaryClass={physics.ao-ph},
      url={https://arxiv.org/abs/2510.02415}, 
}

@article{Beck2018,
  author    = {Beck, Hylke E. and Zimmermann, Niklaus E. and McVicar, Tim R. and Vergopolan, Noemi and Berg, Alexis and Wood, Eric F.},
  title     = {Present and future {K\"o}ppen-Geiger climate classification maps at 1-km resolution},
  journal   = {Scientific Data},
  year      = {2018},
  month     = oct,
  volume    = {5},
  number    = {1},
  pages     = {180214},
  issn      = {2052-4463},
  doi       = {10.1038/sdata.2018.214},
  url       = {https://doi.org/10.1038/sdata.2018.214}
}

@article{Chan2015,
  author    = {Chan, Duo and Wu, Qigang},
  title     = {Significant anthropogenic-induced changes of climate classes since 1950},
  journal   = {Scientific Reports},
  year      = {2015},
  month     = aug,
  volume    = {5},
  number    = {1},
  pages     = {13487},
  issn      = {2045-2322},
  doi       = {10.1038/srep13487},
  url       = {https://doi.org/10.1038/srep13487}
}

@article{Kochkov2024,
  author    = {Kochkov, Dmitrii and Yuval, Janni and Langmore, Ian and Norgaard, Peter and Smith, Jamie and Mooers, Griffin and Kl{\"o}wer, Milan and Lottes, James and Rasp, Stephan and D{\"u}ben, Peter and Hatfield, Sam and Battaglia, Peter and Sanchez-Gonzalez, Alvaro and Willson, Matthew and Brenner, Michael P. and Hoyer, Stephan},
  title     = {Neural general circulation models for weather and climate},
  journal   = {Nature},
  year      = {2024},
  month     = aug,
  volume    = {632},
  number    = {8027},
  pages     = {1060--1066},
  issn      = {1476-4687},
  doi       = {10.1038/s41586-024-07744-y},
  url       = {https://doi.org/10.1038/s41586-024-07744-y}
}

@article{liu2022,
  author       = {Zhuang Liu and
                  Hanzi Mao and
                  Chao{-}Yuan Wu and
                  Christoph Feichtenhofer and
                  Trevor Darrell and
                  Saining Xie},
  title        = {A ConvNet for the 2020s},
  journal      = {CoRR},
  volume       = {abs/2201.03545},
  year         = {2022},
  url          = {https://arxiv.org/abs/2201.03545},
  eprinttype   = {arXiv},
  eprint       = {2201.03545},
  bibsource    = {dblp computer science bibliography, https://dblp.org}
}

@ARTICLE{Karniadakis2021,
  title     = "Physics-informed machine learning",
  author    = "Karniadakis, George Em and Kevrekidis, Ioannis G and Lu, Lu and
               Perdikaris, Paris and Wang, Sifan and Yang, Liu",
  journal   = "Nat. Rev. Phys.",
  publisher = "Springer Science and Business Media LLC",
  volume    =  3,
  number    =  6,
  pages     = "422--440",
  month     =  may,
  year      =  2021,
  url       = {https://www.nature.com/articles/s42254-021-00314-5},
  copyright = "https://www.springernature.com/gp/researchers/text-and-data-mining",
  language  = "en"
}

@article {taylor2012,
      author = "Karl E. Taylor and Ronald J. Stouffer and Gerald A. Meehl",
      title = "An Overview of CMIP5 and the Experiment Design",
      journal = "Bulletin of the American Meteorological Society",
      year = "2012",
      publisher = "American Meteorological Society",
      address = "Boston MA, USA",
      volume = "93",
      number = "4",
      doi = "10.1175/BAMS-D-11-00094.1",
      pages=      "485 - 498",
      url = "https://journals.ametsoc.org/view/journals/bams/93/4/bams-d-11-00094.1.xml"
}

@InProceedings{unet,
author="Ronneberger, Olaf
and Fischer, Philipp
and Brox, Thomas",
editor="Navab, Nassir
and Hornegger, Joachim
and Wells, William M.
and Frangi, Alejandro F.",
title="U-Net: Convolutional Networks for Biomedical Image Segmentation",
booktitle="Medical Image Computing and Computer-Assisted Intervention -- MICCAI 2015",
year="2015",
publisher="Springer International Publishing",
address="Cham",
pages="234--241",
isbn="978-3-319-24574-4"
}

@InProceedings{sfno,
  title = 	 {Spherical {F}ourier Neural Operators: Learning Stable Dynamics on the Sphere},
  author =       {Bonev, Boris and Kurth, Thorsten and Hundt, Christian and Pathak, Jaideep and Baust, Maximilian and Kashinath, Karthik and Anandkumar, Anima},
  booktitle = 	 {Proceedings of the 40th International Conference on Machine Learning},
  pages = 	 {2806--2823},
  year = 	 {2023},
  editor = 	 {Krause, Andreas and Brunskill, Emma and Cho, Kyunghyun and Engelhardt, Barbara and Sabato, Sivan and Scarlett, Jonathan},
  volume = 	 {202},
  series = 	 {Proceedings of Machine Learning Research},
  month = 	 {23--29 Jul},
  publisher =    {PMLR},
  url = 	 {https://proceedings.mlr.press/v202/bonev23a.html}
}

@article{Watt-Meyer2025,
  author    = {Watt-Meyer, Oliver and Henn, Brian and McGibbon, Jeremy and Clark, Spencer K. and Kwa, Anna and Perkins, W. Andre and Wu, Elynn and Harris, Lucas and Bretherton, Christopher S.},
  title     = {{ACE2}: accurately learning subseasonal to decadal atmospheric variability and forced responses},
  journal   = {npj Climate and Atmospheric Science},
  year      = {2025},
  month     = may,
  volume    = {8},
  number    = {1},
  pages     = {205},
  issn      = {2397-3722},
  doi       = {10.1038/s41612-025-01090-0},
  url       = {https://doi.org/10.1038/s41612-025-01090-0}
}

@article{koppen2011,
author = "K{\"o}ppen, Wladimir",
journal = "Meteorologische Zeitschrift",
month = 06,
year = 2011,
title = "The thermal zones of the Earth according to the duration of hot, moderate and cold periods and to the impact of heat on the organic world",
number = "3",
volume = "20",
pages = {351-360},
url = "http://dx.doi.org/10.1127/0941-2948/2011/105",
doi = "10.1127/0941-2948/2011/105",
publisher = "Schweizerbart Science Publishers",
address = "Stuttgart, Germany"
}

@misc{cBottle,
      title={Climate in a Bottle: Towards a Generative Foundation Model for the Kilometer-Scale Global Atmosphere}, 
      author={Noah D. Brenowitz and Tao Ge and Akshay Subramaniam and Peter Manshausen and Aayush Gupta and David M. Hall and Morteza Mardani and Arash Vahdat and Karthik Kashinath and Michael S. Pritchard},
      year={2025},
      eprint={2505.06474},
      archivePrefix={arXiv},
      primaryClass={physics.ao-ph},
      url={https://arxiv.org/abs/2505.06474}, 
}

@article{dlesym,
author = {Cresswell-Clay, Nathaniel and Liu, Bowen and Durran, Dale R. and Liu, Zihui and Espinosa, Zachary I. and Moreno, Raul A. and Karlbauer, Matthias},
title = {A Deep Learning Earth System Model for Efficient Simulation of the Observed Climate},
journal = {AGU Advances},
volume = {6},
number = {4},
pages = {e2025AV001706},
doi = {https://doi.org/10.1029/2025AV001706},
url = {https://agupubs.onlinelibrary.wiley.com/doi/abs/10.1029/2025AV001706},
eprint = {https://agupubs.onlinelibrary.wiley.com/doi/pdf/10.1029/2025AV001706},
note = {e2025AV001706 2025AV001706},
year = {2025}
}
% don't specify bibliographystyle
%
%%%%%%%%%%%%%%%%%%%%%%%%%%%%%%%%%%%%%%%%%%%%%%%
\clearpage
\appendix
\begingroup
\setlength{\intextsep}{0pt}
\section{}

\begin{figure}[H]
\centering
\noindent\includegraphics[width=0.8\textwidth]{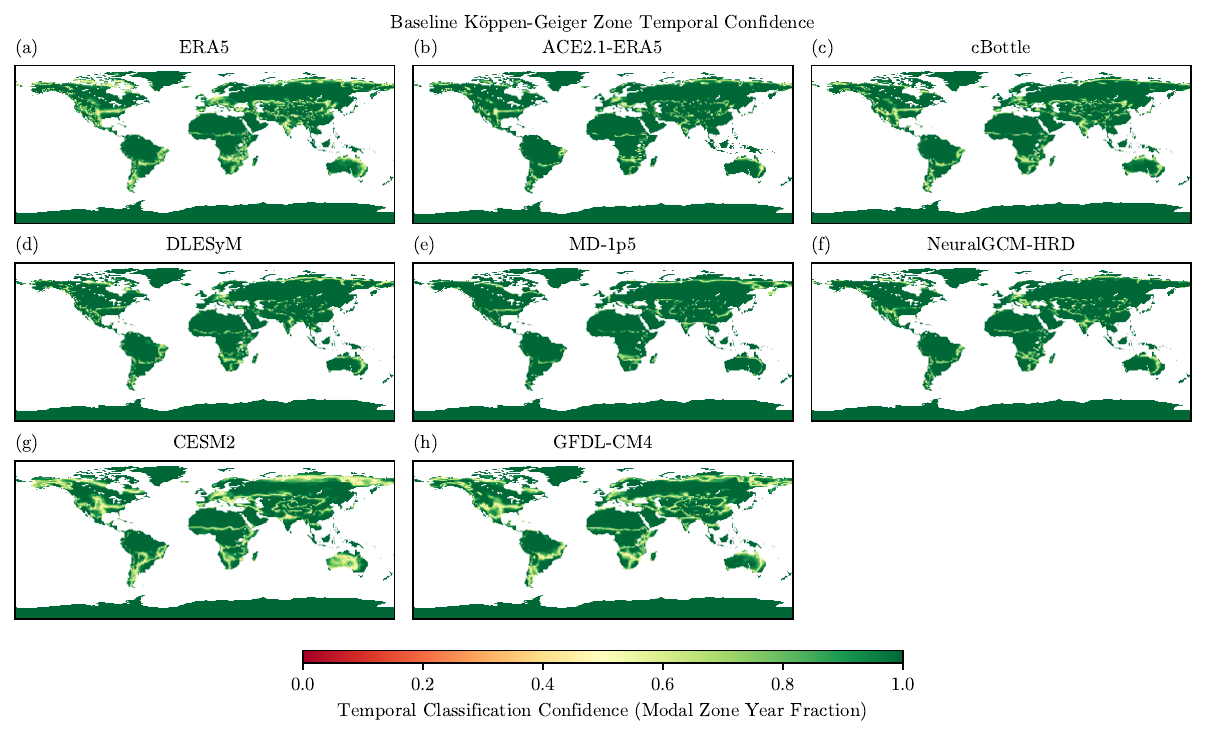}
\caption{\textbf{Temporal Confidence of Baseline K{\"o}ppen-Geiger Classification.} For each land cell, we show the fraction of baseline years (1979-2014) whose single-year major-zone classification matches the cell's modal zone for ERA5 (a), AIMIP models (b), and the two reference AMIP models (g-h), computed from their ensemble mean. Values near 1 (green) indicate climate zone stability across interannual variability. Lower values near 0 (red) mark transition regions where the assigned zone is sensitive.}
\label{figs2}
\end{figure}

\begin{figure}[H]
\centering
\noindent\includegraphics[width=0.8\textwidth]{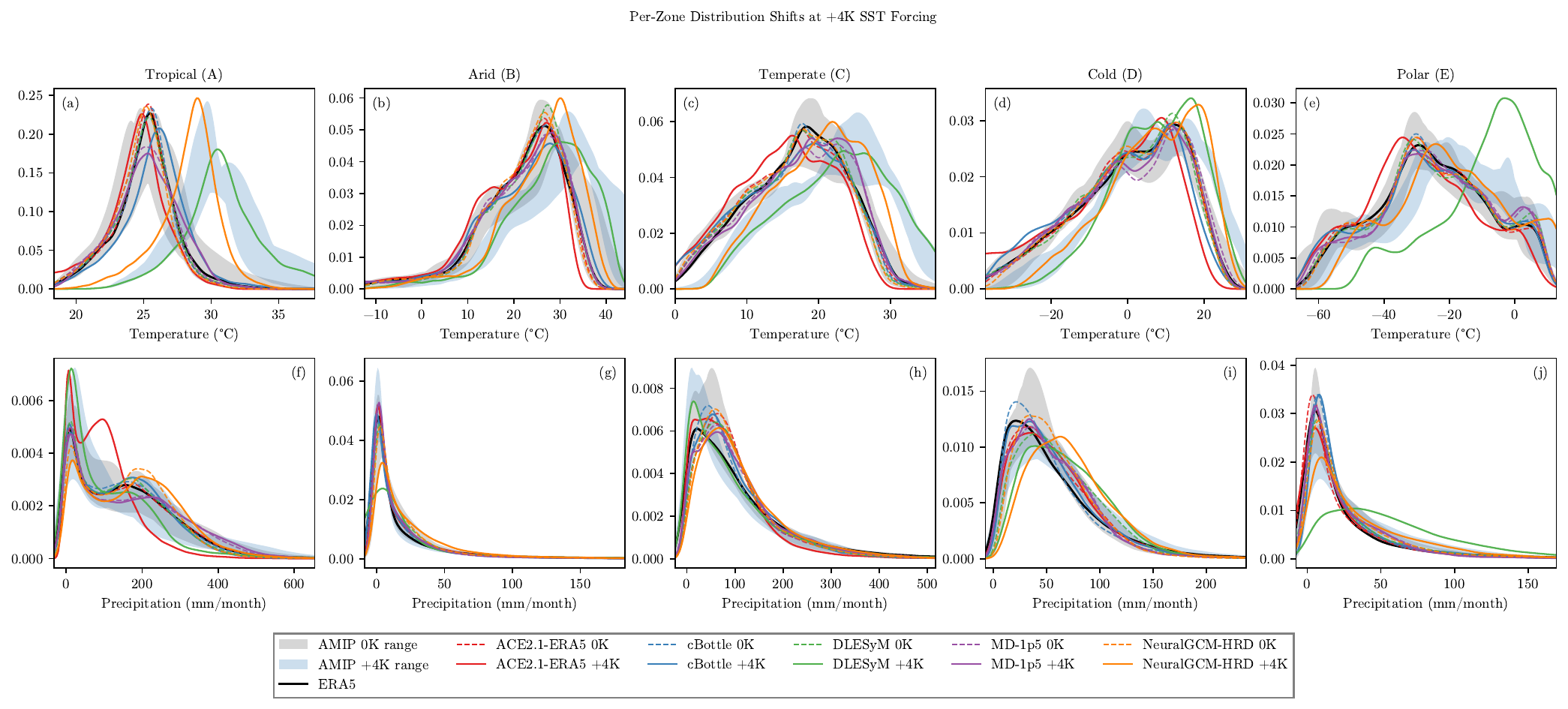}
\caption{\textbf{Per-Zone Temperature and Precipitation Distribution Shifts at +4\,K SST Forcing.} We present the kernel density estimates of monthly temperature (first row; a-e; \textdegree\, C) and precipitation (second row; f-j; mm\, month$^{-1}$) pooled over the baseline period (1979-2014) and weighted by cosine-latitude area for each K{\"o}ppen-Geiger major zone (columns). We fix zone membership at each model's 0\,K modal classification and apply to the 0\,K (dashed) and +4\,K (solid) fields. ERA5 (black) uses its own baseline classification. Shading spans the min-max envelope of the AMIP ensemble at 0,K (grey; 22 models) and +4\,K (blue; 13 models).}
\label{figs1}
\end{figure}

\begin{table}[H]
\caption{\textbf{Relevant CMIP6 AMIP Models.} All fields are regridded bilinearly to a $1^\circ\times1^\circ$ grid over 1979-2014 (36 years) and ensemble-averaged per scenario. The Members (amip) column counts the number of observed-SST control members averaged into each model's 0K field. The perturbation experiments p4K and m4K (uniform $+4$ and $-4$\,K SST forcing, CMIP6 amip-p4k and amip-m4k) each provide a single member (two for NorESM2-LM). The perturbation subset (13 models with $+4$\,K and 9 with $-4$\,K) forms the AMIP envelope, while all 22 models are included in the 0K baseline multi-model mean. Near-surface specific humidity (huss) is only relevant for Figure 3k-o. The Figures column lists the figures that included the model. CESM2 and GFDL-CM4 are shown individually as named references. CESM2's amip-p4K run ends December 2013, so we trim its control baseline to 35 years.}
\label{tab:amip_models}
\centering
\begin{tabular}{l c c c c}
\hline
 \textbf{Model} & \textbf{Members (amip)} & \textbf{SST scenarios} & \textbf{huss} & \textbf{Figures} \\
\hline
\multicolumn{5}{l}{\textit{Perturbation subset (control \& uniform-SST forcing)}} \\
TaiESM1 & 1  & amip, p4K, m4K & Y & 01-04 \\
BCC-CSM2-MR & 1  & amip, p4K, m4K & N & 01-04 \\
CanESM5 & 7  & amip, p4K, m4K & Y & 01-04 \\
CNRM-CM6-1 & 1  & amip, p4K, m4K & Y & 01-04 \\
IPSL-CM6A-LR & 1  & amip, p4K, m4K & Y & 01-04 \\
HadGEM3-GC31-LL & 5  & amip, p4K, m4K & Y & 01-04 \\
MRI-ESM2-0 & 1  & amip, p4K, m4K & Y & 01-04 \\
CESM2 & 1  & amip, p4K, m4K & Y & 01-04 \\
GFDL-CM4 & 1  & amip, p4K, m4K & Y & 01-04 \\
E3SM-1-0 & 1  & amip, p4K & Y & 01-04 \\
MIROC6 & 9  & amip, p4K & Y & 01-04 \\
GISS-E2-1-G & 20 & amip, p4K & Y & 01-04 \\
NorESM2-LM & 2  & amip, p4K & Y & 01-04 \\
\hline
\multicolumn{5}{l}{\textit{Baseline-only (control SST; 0K multi-model mean)}} \\
ACCESS-CM2 & 7  & amip & - & 01 \\
ACCESS-ESM1-5 & 9  & amip & - & 01 \\
EC-Earth3 & 5  & amip & - & 01 \\
EC-Earth3-CC & 5  & amip & - & 01 \\
FIO-ESM-2-0 & 6  & amip & - & 01 \\
IPSL-CM6A-MR1 & 9  & amip & - & 01 \\
GISS-E2-2-G & 5  & amip & - & 01 \\
NorCPM1 & 9  & amip & - & 01 \\
NESM3 & 5  & amip & - & 01 \\
\hline
\end{tabular}
\end{table}

\begin{table}[h]
\caption{\textbf{Relevant AIMIP Phase 1 Submissions.} These five models provide the minimum variable set for the K\"oppen--Geiger classification. All outputs are regridded to a $1^\circ\times1^\circ$ grid over 1979-2014. Each model contributes a five-member baseline ensemble. All models also contribute at least one member for the uniform $+2$\,K and $+4$\,K SST-warming experiments (aimip-p2k and aimip-p4k). Near-surface specific humidity (huss) is an output for ACE2.1-ERA5 and MD-1p5, derived from surface pressure and 2m dewpoint for cBottle-1.3 and NeuralGCM-HRD, and unavailable for DLESyM. Training periods are 1979-2014 (ERA5) for all models except DLESyM (1983-2016, ERA5 plus satellite observations), which overlaps with the AIMIP 2015-2024 holdout.}
\label{tab:aimip_models}
\centering
\begin{tabular}{>{\centering\arraybackslash}p{0.28\textwidth}>{\centering\arraybackslash}p{0.23\textwidth}>{\centering\arraybackslash}p{0.13\textwidth}> {\centering\arraybackslash}p{0.13\textwidth}>{\centering\arraybackslash}p{0.13\textwidth}}
\hline
\textbf{Model} & \textbf{Architecture} & \textbf{Grid} & \textbf{Training} & \textbf{huss} \\
\hline
ACE2.1-ERA5 \newline \cite{Watt-Meyer2025} & Data-driven autoregressive \newline (SFNO) & $1^\circ$ & 1979-2014 & Y \\
DLESyM \newline \cite{cresswell} & Data-driven autoregressive \newline(ConvNeXt U-Net \& GRU) & $1^\circ$ & 1983-2016 & N \\
NeuralGCM-HRD \newline \cite{Kochkov2024,yuval2026} & Hybrid physics-AI \newline (differentiable dynamical core \& learned parametrizations) & $1^\circ$\newline (from $2.8^\circ$) & 1979-2014 & Y (derived) \\
cBottle-1.3 \newline \cite{cBottle} & Conditional generative diffusion \newline (two-stage denoising) & HEALPix~6 ($\sim1^\circ$) & 1979-2014 & Y (derived) \\
MD-1p5 \newline \cite{hall2026} & Autoregressive latent diffusion \newline (encoder-DDPM-decoder) & $1^\circ$ & 1979-2014 & Y \\
\hline
\end{tabular}
\end{table}
\endgroup
%\bibliography{ enter your bibtex bibliography filename here }

%Reference citation instructions and examples:
%
% Please use ONLY \cite and \citeA for reference citations.
% \cite for parenthetical references
% ...as shown in recent studies (Simpson et al., 2019)
% \citeA for in-text citations
% ...Simpson et al. (2019) have shown...
%
%
%...as shown by \citeA{jskilby}.
%...as shown by \citeA{lewin76}, \citeA{carson86}, \citeA{bartoldy02}, and \citeA{rinaldi03}.
%...has been shown \cite{jskilbye}.
%...has been shown \cite{lewin76,carson86,bartoldy02,rinaldi03}.
%... \cite <i.e.>[]{lewin76,carson86,bartoldy02,rinaldi03}.
%...has been shown by \cite <e.g.,>[and others]{lewin76}.
%
% apacite uses < > for prenotes and [ ] for postnotes
% DO NOT use other cite commands (e.g., \citet, \citep, \citeyear, \nocite, \citealp, etc.).
%

\end{document}